\documentstyle[aps,prl,multicol]{revtex}  
\begin{document}
\title{Gaussian density fluctuations and Mode Coupling Theory for 
supercooled liquids}
\author{E. Zaccarelli$^1$, G. Foffi$^1$, F. Sciortino$^2$, 
P. Tartaglia$^2$ and K. A. Dawson$^1$}
\address{$^1$ Irish Centre for Colloid Science and Biomaterials, Department of
Chemistry, \\ 
University College Dublin, Belfield, Dublin 4, Ireland \\ 
$^2$ Dipartimento di Fisica, Universit\`{a} di Roma La Sapienza\\
and  Istituto Nazionale di Fisica della Materia, Unit\`{a} di Roma La Sapienza
\\ Piazzale Aldo Moro 2, 00185 Roma, Italy}
\date{\today}
\maketitle

\begin{abstract}
The equations of motion for the density modes of a fluid, derived from
Newton's equations, are written as a linear generalized Langevin equation.
The constraint imposed by the fluctuation-dissipation theorem is
used to derive an exact form for the memory function. 
The resulting equations, solved under the assumption that the
noise, and consequently density fluctuations, of the liquid are gaussian distrib
uted,
are equivalent to the random-phase-approximation for the 
static structure factor and to the well known ideal mode 
coupling theory (MCT) equations for the dynamics.
This finding suggests that MCT is the canonical mean-field theory of the 
fluid dynamics.
\end{abstract}


\begin{multicols}{2}   

The  description of the  
dynamics of supercooled liquids is one of the most intriguing goals 
in  condensed-matter physics. 
Different approaches have been pursued in the last decades
\cite{gotzesjolander,mazenko,smith,kawasaki,thirumalai,coniglio,parisi}, 
originating both from the physics of liquids
and from the physics of disordered systems.
Many of these approaches strongly suggest 
that two different mechanisms for the decay of fluctuations are active in
 two different temperature ranges. A cross-over temperature $T_c$ separates 
the region of 'weak' supercooling from the region of 'strong' supercooling.

The description of the long-time dynamics and the associated 
evaluation of the transport coefficient in the strong supercooling region 
(below $T_c$) has proved  
an extremely difficult task\cite{mazenko,kawasaki,extendedMCT}.
The 
thermodynamic description of the liquid state between $T_c$ and the
calorimetric glass transition has been attempted, but no 
well defined connection between thermodynamics and dynamics has been
achieved as yet. In the weak supercooling regime, detailed predictions for the
 space and 
time dependence of the long time  decay of density correlations
have been formulated using the ideal mode coupling theory 
(MCT)\cite{gotzesjolander}, one
of the first approaches to  identify the existence of 
the cross-over temperature.
The agreement of
MCT predictions with experimental findings\cite{vanmegen,goetzereview}
and {\em exact} correlation functions
evaluated from  molecular dynamics 
simulations\cite{kobprlgleim,skcondmat,fabbian} --- both 
for atomic and molecular models --- supports the view that MCT is indeed  
able to describe the slow dynamics in weak supercooled states. 
 
Despite its remarkable practical success, the 
presence of apparently uncontrolled approximations 
in the derivation of the MCT equations makes it difficult to gain insights 
into possible improvements of the theory.
The aim of this Letter 
is to present a new derivation of the ideal MCT equations, 
starting from the microscopic equations for the evolution of the
density (Newton's equations) and writing them as a linear generalized 
Langevin equation.
A formally exact expression for the memory kernel is derived and, on making 
the approximation that the noise in the Langevin equation is Gaussian, 
the standard MCT equations are obtained.
Note that the proposition of Gaussian noise implies that  the density 
fluctuations are also gaussian\cite{pratt}. 
The outcome of the present approach offers  a 
route to MCT that involves clear 
and known approximations for static 
and dynamics. At the same time it suggests how MCT equations 
can be developed for more complex systems such as molecules and polymers.   

The density of a system composed of  
$N$ particles located at positions ${\bf r}_j$ 
is  $\rho({\bf r},t) \equiv \sum_{j=1}^N \delta({\bf r}-{\bf r}_j(t))$.
The Fourier transform of   $\rho({\bf r},t)$  is  
$\rho_{\bf k}(t) \equiv \sum_{j=1}^N e^{i{\bf k}\cdot {\bf r}_j(t)}$.
In the case of a pairwise additive potential, 
the time evolution of $\rho_{\bf k}(t)$ can be written as

\begin{eqnarray} 
\ddot{\rho}_{\bf k}(t)&=&
-\sum_j({\bf k}\cdot {\bf \dot{r}}_j(t))^2 e^{i{\bf k}\cdot 
{\bf r}_j(t)}-\nonumber\\&-&
\frac{1}{mV}\sum_{\bf k'}v_{k'} ({\bf k}\cdot {\bf k'}) \rho_{{\bf k}-{\bf
k'}}(t)\rho_{\bf k'}(t)
\label{eq:newton}
\end{eqnarray}

\noindent 
where $v_k$ is the Fourier transform of the pair potential. 
Eq.\ref{eq:newton}
is the Newton equation for the variables $\rho_{\bf k}(t)$.

The equation of motion for the density (Eq.~\ref{eq:newton}) 
can be rewritten as,

\begin{equation} 
\ddot{\rho}_{\bf k}(t)+\omega^2_{k}\rho_{\bf k}(t)=
{\cal F}_{\bf k}(t) 
\label{eq:trialeq}
\end{equation}

\noindent
where the  linear term in $\rho_{\bf k}(t)$, that represents 
the elementary excitations of the system, has been explicitly isolated.
These excitations can be interpreted as `phonons',
that would oscillate with frequencies  
$\omega^2_{k}$ if no interaction force amongst them were present.
Combining Eqs.\ref{eq:newton} and \ref{eq:trialeq} the residual 
force ${\cal F}_{\bf k}(t)$ has the formal expression

\begin{eqnarray} 
{\cal F}_{\bf k}(t)&=&{\omega}^2_k\rho_{\bf k}(t)-
\sum_j 
({\bf k}\cdot {\bf \dot{r}}_j(t))^2 e^{i{\bf k}\cdot {\bf r}_j(t)}
-\nonumber\\
&-&\frac{1}{mV}\sum_{\bf k'}v_{k'} ({\bf k}\cdot {\bf k'}) 
\rho_{{\bf k}-{\bf
k'}}(t)\rho_{\bf k'}(t) 
\label{eq:residual}
\end{eqnarray}

\noindent
The value of $\omega^2_{k}$ 
that  results in the least residual interaction between phonons, 
i.e. the one for which $\partial \langle|{\cal F}_{\bf k}|^2\rangle / 
\partial \omega^2_{k}=0$ is \cite{zwanzig,percus}

\begin{equation}
\omega^2_{k} =\frac{k^2}{\beta m S_k}.
\label{eq:w2}
\end{equation}

\noindent
This choice for $\omega^2_{k}$ is also imposed by requiring the 
correct short-time  limit of Eq.\ref{eq:newton}\cite{hansen}.
The minimization principle produces an
orthogonality condition between the force and the density variables 
at all times,

\begin{equation} 
\langle \rho_{-{\bf k}}(t)  {\cal F}_{\bf k}(t) \rangle =0
\label{eq:ortg}
\end{equation}

\noindent
Substituting the expression for $ {\cal F}_{\bf k}(t)$ (Eq.~\ref{eq:residual})
in this orthogonality condition leads to 
the exact Yvon-Born-Green (YBG)
equation\cite{tosi}

\begin{equation} 
S_k= 1- \frac{\beta}{N k^2 V}
\sum_{\bf k'}v_{\bf k'} ({\bf k}\cdot {\bf k'}) \langle \rho_{-{\bf
k}}(t) \rho_{{\bf k}-{\bf k'}}(t)\rho_{\bf k'}(t)\rangle 
\label{eq:ybg}
\end{equation}

\noindent
where  $S_k \equiv \langle 
\rho_{-{\bf k}}\rho_{{\bf k}}\rangle / N $ is the static structure factor,
$\beta \equiv 1/k_BT$, $k_B$ 
is the Boltzmann constant, $T$ is the temperature and
the symbols $\langle \cdot \cdot \cdot \rangle$ indicate equilibrium averages.

Following the spirit of the 
Zwanzig-Mori (ZM) formalism \cite{mori}, the residual force can be 
written as sum of two contributions: (i)
a term of dissipative origin, containing a memory
function dependent on time. This represents slow deviations of the system 
from the integrable parts of phase space. (ii) 
A general random noise term that represents a more `chaotic' 
behavior for the system. Hence,

\begin{equation} 
{\cal F}_{\bf k}(t)=-\int_0^t \gamma_{\bf k}(t-t') 
\partial_{t'} \rho_{\bf
k}(t') dt' + f_{\bf k}(t) 
\label{eq:expansion}
\end{equation}

\noindent
where $\gamma_{\bf k}(t)$ acts as memory function of the 
system and $f_{\bf k}(t)$ as fluctuating force. Note that
the previous 
choice for the residual force  (Eq.~\ref{eq:expansion}) 
satisfies the orthogonality condition,  Eq.~\ref{eq:ortg}.

The functions  $f_{\bf k}(t)$ and $\gamma_{\bf k}(t)$ are not independent. 
Indeed, they must satisfy 
the fluctuation-dissipation theorem (FDT)
to guarantee that the long time evolution of the system is consistent with 
the  correct Boltzmann equilibrium distribution. This requires that
the autocorrelation function of $f_{\bf k}(t)$ is proportional to
the memory function $\gamma_{\bf k}(t)$. Also,
the average over the noise is zero. Thus,

\begin{equation}
 \langle f_{\bf k}(t)\rangle=0
\label{eq:fnorm} 
\end{equation}
\begin{equation}
 \frac{\langle f_{-{\bf k}}(t)f_{\bf k}(t')\rangle}
{\langle|\dot{\rho}_{\bf
k}|^2\rangle}= \gamma_{\bf k}(t-t')
\label{eq:fdt}
\end{equation}

\noindent
As a  result of all these formal assignments 
the Newton's equations for the density fluctuations are 
rewritten in a form similar to
a linear generalized Langevin equation as,

\begin{equation} 
\ddot{\rho}_{\bf k}(t)+{\omega}^{2 \ }_k\rho_{\bf k}(t)+
\int_0^t \gamma_{\bf k}(t-t') \partial_{t'} \rho_{\bf
k}(t') dt' = f_{\bf k}(t) 
\label{eq:langevin}
\end{equation}

\noindent

Note that the transition from the Newton's equations to the generalized
Langevin equation is associated with a transition from
averages over the initial conditions to averages over
the realization of the noise (defined in Eqs.~\ref{eq:fnorm}-
\ref{eq:langevin}). FDT
guarantees that the dynamics generated by
this stochastic process leads to thermodynamic equilibrium. 

Using  Eq. \ref{eq:fdt}, we calculate
the explicit exact form for the memory function\cite{comecalcologamma} 
as, 

\begin{eqnarray}
&&\gamma_{\bf k}(t)=\frac{\beta m}{N k^2} \left[
\left(\frac{k^2}{\beta m}\right)^2(n^2c_k^2-1)NS_k(t)+\right.
\nonumber\\ &&+\left.
\langle \sum_l\sum_m ({\bf k}\cdot {\bf \dot{r}}_l(0))^2 e^{-i{\bf k}\cdot 
{\bf r}_l(0)} ({\bf k}\cdot {\bf \dot{r}}_m(t))^2 e^{i{\bf k}\cdot 
{\bf r}_m(t)} \rangle +\right.
\nonumber\\ &&+\left.
\frac{1}{(m V)^2}
\sum_{\bf k'}\sum_{\bf k''}v_{k'}v_{k''}({\bf k}\cdot {\bf k'})
(-{\bf k}\cdot {\bf k''})\right.
\nonumber\\ &&\left.
\langle \rho_{-{\bf k}-{\bf k''}}(0) \rho_{\bf k''}(0) 
\rho_{{\bf k}-{\bf k'}}(t) \rho_{\bf k'}(t)\rangle \right.
+\nonumber\\ &&+\left.
\frac{n k^2}{\beta m^2 V} c_k \left\{
\sum_{\bf k'} v_{k'}({\bf k}\cdot {\bf k'})\langle \rho_{-{\bf k}}(0)
\rho_{{\bf k}-{\bf k'}}(t) \rho_{\bf k'}(t)\rangle +\right.\right.
\nonumber\\ && \left.\left.
+\sum_{\bf k''} v_{k''}(-{\bf k}\cdot {\bf k''})\langle
\rho_{-{\bf k}-{\bf k''}}(0) \rho_{\bf k''}(0)\rho_{\bf k}(t) \rangle \right\}
\right.
-\nonumber\\ &&-\left.
\frac{ k^2}{\beta m}n N c_k \int_{0}^t dt' 
\gamma_{\bf k}(t-t')\partial_{t'}S_k(t') 
-\frac{1}{mV}\sum_{\bf k'} v_{k'}\right.\nonumber\\ &&\left.
(-{\bf k}\cdot {\bf k'})\int_{0}^t dt' 
\gamma_{\bf k}(t-t')
\langle \rho_{-{\bf k}-{\bf k'}}(0) \rho_{\bf k'}(0)\dot{\rho}_{\bf k}(t')
\rangle \right].
\label{eq:gamma}
\end{eqnarray}

Eqs.~\ref{eq:fnorm}-\ref{eq:gamma}
constitute an exact definition of the noise process and have been obtained
by using only FDT and the causality
relation $\langle \rho_{-{\bf k}}(0)f_{\bf k}(t)\rangle=0$.
These equations have the same content as the equations derived
in the ZM formalism\cite{mori}.
To explicitly evaluate the $\gamma_{\bf k}(t)$ function, 
approximations have to be made on the 
stochastic processes $\rho_{\bf k}(t)$ and $f_{{\bf k}}(t)$,
i.e. the density and the noise \cite{note}.

In this Letter, a simple approximation is considered.
We assume that 
$f_{\bf k}(t)$ is an additive gaussian process. 
This is the major approximation, since  the assumption of
a gaussian process and the linearity of Eq.~\ref{eq:langevin} imply that
$\rho_{\bf k}(t)$ is also a gaussian process. 
As shown in Eqs.~\ref{eq:residual} and \ref{eq:expansion} 
the true $f_{\bf k}(t)$ 
is a non linear function of $\rho_{\bf k}(t)$. 
This  indicates that in an exact theory  
both quantities can not be simultaneously gaussian distributed
\cite{kawasaki}.

%

A less significant approximation consists in 
averaging over the velocities  
neglecting the fluctuations in the
single particle kinetic energy, i.e. assuming

\begin{eqnarray}
\langle \sum_l\sum_m ({\bf k}\cdot {\bf \dot{r}}_l(0))^2 e^{-i{\bf k}\cdot 
{\bf r}_l(0)} ({\bf k}\cdot {\bf \dot{r}}_m(t))^2 e^{i{\bf k}\cdot 
{\bf r}_m(t)} \rangle \nonumber\\
\approx \frac{k^4}{\beta^2 m^2} N S_k(t) 
\label{eq:approxv}
\end{eqnarray}

\noindent
The gaussian nature of the noise ensures certain 
simplifications in the properties of  $\rho_{\bf k}(t)$ 
 and this allows to calculate the
 multiple averages in Eq.~\ref{eq:gamma}.
Together with the approximation in Eq.~\ref{eq:approxv}, this allows to
 derive an expression for $\gamma_{\bf k}(t)$ which requires as 
input the density-density 
correlation functions. As a result, the gaussian approximation in the density
gives,

\begin{eqnarray}
\!\!\!\!\!&&\gamma_{\bf k}(t)=\frac{n \beta}{m V k^2}
\sum_{{\bf k'}\neq {\bf k}}\{v_{k'}^2({\bf k}\cdot {\bf k'})^2
+ v_{k'}v_{k-k'}({\bf k}\cdot {\bf k'})\nonumber \\
&&({\bf k}\cdot ({\bf k} -{\bf k'}))
\} S_{|{\bf k}-{\bf k'}|}(t) S_{k'}(t) 
+ \frac{k^2 n^2}{\beta m }(c_k+\beta v_k)^2 S_k(t)- \nonumber \\
&&-n (c_k+\beta v_k)\int_{0}^t dt' 
\gamma_{\bf k}(t-t')\partial_{t'}S_k(t') 
\label{eq:resultgamma}
\end{eqnarray}

\noindent
The previous expression still contains the Fourier transform of the
pair potential $v_k$.  As is well known\cite{hansen,tosi}, once
the gaussian approximation for the density
has been made, the YBG equation (Eq.~\ref{eq:ybg}) 
can be consistently solved\cite{ybg-gaussian}, providing the so-called 
Random-Phase-Approximation (RPA)\cite{hansen,pines}, $c_k=-\beta
v_k$, between the direct correlation function 
$c_k \equiv (1-1/S(k))/ n$ and the potential $v_k$.
Eliminating in Eq.\ref{eq:resultgamma} $v_k$ in favor of $c_k$
one obtains,

\begin{eqnarray} &&\gamma_{\bf k}(t)^{^{MCT}} = \frac{n}{\beta m  k^2 V}
\sum_{{\bf k'}\neq {\bf k}}\{({\bf k}\cdot {\bf k'})^2
c_{\bf k'}^2+ 
\nonumber\\ &+&({\bf k}\cdot {\bf k'})({\bf k}\cdot ({\bf k} -{\bf k'}))
c_{\bf k'}
c_{{\bf k} -{\bf k'}}\} S(|{\bf k}-{\bf k'}|,t) S(k',t).
\label{eq:memoryRPA}
\end{eqnarray}

$\gamma_{\bf k}(t)^{^{MCT}}$ coincides with the memory function
that has been originally calculated in Ref.~\cite{gotzesjolander}
within the ideal MCT framework.

The reader may note that the  choice of  a gaussian process 
for $\rho_{\bf k}(t)$, if used 
directly in Eq.\ref{eq:trialeq}-\ref{eq:residual}, implies that 
the time evolution of the 
density-density correlation function is described by   
undamped harmonic modes\cite{pines,spiegazione} with frequency
$\omega_k$. In this approximation 
there is no interaction between the modes and hence 
strictly speaking $\gamma_{\bf k}(t)=0$. In the present derivation,
the memory kernel is not zero, since the assumption that the noise is 
gaussian is 
introduced  only after the residual interactions are constrained to be
a noise and dissipation and a formally correct expression 
for $\gamma_{\bf k}(t)$ (Eq.~\ref{eq:gamma}) is derived.
This apparent contradiction is a consequence of
the fact that the same approximations made before constraining
phase space to have the Mori-type properties are not equivalent to those
made after this fundamental constraint is forced upon the system. 
Therefore, the division of the 
residual forces between density waves into a dissipative and noise term
is fundamental. 
This step, whilst not always acceptable, is a 
conventional approach to dealing with 
complex many body forces, and the fact that the integrable  motions have 
been clearly removed from the interactions prior to this division is quite 
satisfying.

The present calculations show the following.

(i) The assumption of
gaussian properties for $\rho_{\bf k}(t)$ and  $f_{\bf k}(t)$ 
allows to derive  MCT. All approximations used in the
conventional MCT are exact in this limit.
Hence MCT can be seen, as a minimum,  as an exact theory in the RPA limit and
a fully consistent 
mean-field approximation to the dynamics of a complex system.
Thus, the MCT dynamics is in the same class of 
universality as mean-field dynamics.  
This possibility was suggested some time ago on the basis of the   
analogies between the equations describing the 
schematic MCT models and the 
the dynamics of the order parameter in 
disordered $p$-spin models, solved under strict 
mean-field approximation\cite{thirumalai,crisanti}.

(ii) The present approach offers a more direct interpretation to the
uncontrolled approximation intrinsic in the conventional MCT. 
Indeed, as clearly stated in Ref.\cite{gotzesjolander}, 
{\it a priori} 
estimation of the quality of the conventional MCT are not known. 
The derivation of the 
basic equations does not include any
systematic expansion scheme.
In the present approach, the more transparent approximations 
suggest that improvement over the MCT predictions is
feasible. In particular, it is possible to eliminate $v_k$
in Eq.\ref{eq:gamma} by implementing higher order approximations
for the triplet correlation function. For example, it is possible
to implement in the present scheme the 
Singwi-Sj\"{o}lander (STLS) closure\cite{ss}, which introduces controlled
corrections to the gaussian statistics. Work is in progress on this
topic.

(iii) In the RPA, the equation for the memory kernel, Eq. 
\ref{eq:gamma}, simplifies and the
integral contributions cancel out. This simplification does not
occur in general. It may be shown that\cite{expanded} 
these terms provide a better
description of the short time dynamics as well as a 
renormalization of the memory at infinite time. It can also be shown that
these terms are  dropped in the conventional MCT, 
when projecting onto the density pair 
subspace. In the present approach, these terms can be 
added to the conventional MCT result. This 
will provide a consistent expression for
the short time memory function.

(iv) An interesting aspect of the G\"otze MCT derivation is that it
does not enforce the consistency conditions between static and
dynamics and therefore the input structure factors do not have to be
RPA in origin (and indeed, ``exact'' direct correlation functions
calculated from simulation data are often used as input in the
theoretical calculations\cite{kobprlgleim,fabbian}).  In the
conventional MCT approach, $v_k$ disappears because of the gaussian
approximation in calculating the normalization matrix of the
projection\cite{gotzesjolander}. 
If consistency between statics and dynamics were enforced,
the RPA would have followed, as discussed above. Moreover,
Kawasaki\cite{kawasaki} has recently presented a derivation of MCT
based on the (quadratic) density functional Ramakrishnan-Yussouf free
energy of a liquid, where the effective interaction between the
density pairs is exactly $- c(k) / \beta$, suggesting that it is
possible to correct the theory, without fundamentally affecting the
universality class of its dynamics.  Perhaps the answer to the
intriguing question of why this is successful lies in formulating
higher corrections to the YBG and memory kernel equations.

(v) Finally, we stress that the present calculation will  
open new possibilities for an improved MCT description of the
dynamics of deep supercooled liquids  as well as 
extensions in order to deal with more complex systems, like
molecules\cite{fabbian,mctmolecole,mctsiti,franoschdumbell} 
and polymers\cite{mctpolymers}.
This is a strategy actively pursued, 
and it will be most interesting to see the outcome.

We thank A. Crisanti, S. Ciuchi, W. G\"otze, K. Kawasaki 
for discussions and suggestions. F.S. and P.T. are supported by
INFM-PRA-HOP and MURST PRIN98, E.Z., G.F. and K.D. by COST P1.

\end{multicols} 

\begin{references}

\bibitem{gotzesjolander} U. Bengtzelius, W. G\"{o}tze and A. Sj\"olander,
J. Phys. C {\bf 17}, 5915 (1984); 
W. G\"{o}tze, in Liquids, Freezing and Glass Transition, 
J.P. Hansen, D.Levesque and J.Zinn-Justin Eds.,
(North-Holland, Amsterdam, 1991).

\bibitem{mazenko} S. P. Das and G. F. Mazenko, Phys. Rev. A {\bf 34}, 2265 
(1986).

\bibitem{smith} J. W. Dufty and R. Schmitz, Transp. Theory Stat. Phys. 
{\bf 24}, 903 (1995) 


\bibitem{kawasaki} K. Kawasaki, Transp. Theory Stat. Phys. {\bf 24}, 755 
(1995); Physica A {\bf 208}, 35 (1994).

\bibitem{thirumalai} T. R. Kirkpatrick and D. Thirumalai,
Phys. Rev. Lett.  {\bf 58}, 2091 (1987); 



\bibitem{coniglio} A.Coniglio, A. de Candia, A. Fierro and M. Nicodemi
J. Phys. Cond. Matter {\bf 11}, A167 (1999).


\bibitem{parisi} M. M\'ezard and G. Parisi, 
J. Phys. Cond.  Matter {\bf 11}, A157 (1999);
M. Cardenas, S. Franz and G. Parisi,  J. Chem. Phys. {\bf 110}, 1726 
(1999).


\bibitem{extendedMCT} L. Sj\"ogren, Phys. Rev. A {\bf 22}, 2866 (1980);
W. G\"{o}tze and L. Sj\"ogren, Z. Phys. B {\bf 65}, 415 (1987).

\bibitem{vanmegen} 
W. van Megen, S. M. Underwood and P. N. Pusey, Phys. Rev. Lett. {\bf 67}, 1586 (
1991);
W. van Megen and S. M. Underwood, Phys. Rev. Lett. {\bf 70}, 2766 (1993); 
ibid {\bf 72}, 1773 (1994).

\bibitem{goetzereview} W. G\"{o}tze, J. Phys. Cond. Matt. {\bf 11}, A1 (1999).

\bibitem{kobprlgleim} T. Gleim, W. Kob, K. Binder, Phys. Rev. Lett. 
{\bf 81}, 4404 (1998).

\bibitem{skcondmat} F. Sciortino and W. Kob, Phys. Rev. Lett. in press (2001).


\bibitem{fabbian}
L. Fabbian, A. Latz, R. Schilling, F. Sciortino, P. Tartaglia and C. Theis,
Phys. Rev. E.  {\bf 60}  5768 (1999).




\bibitem{pratt} 
See e.g. D. Chandler, Phys. Rev. E. {\bf 48} 2898 (1993);
G. Hummer, S. Garde, A. E. Garcia, A. Pohorille and L. R. 
Pratt, Proc. Natl. Acad. Sci. {\bf 93}, 8951 (1996).
G. E. Crooks and D. Chandler, Phys. Rev. E {\bf 56} 4217 (1997).
Note that static density fluctuations do not necessarily imply 
Gaussian dynamics.



\bibitem{zwanzig} R. Zwanzig, Phys. Rev. {\bf 156}, 190 (1967).


\bibitem{percus} J. K. Percus, in Studies in Statistical Mechanics, 
Vol. VIII, edited by E. W. Montroll and J.L. Lebowitz (North-Holland, Amsterdam, 1982).


\bibitem{hansen} J. P. Hansen and I. R. McDonald, Theory of simple liquids
(Academic Press, London, 1986), 2nd Edition.

\bibitem{tosi} A.N.H.March and M.P.Tosi, Atomic dynamics in liquids, 
Dover (NY) 1991. 




\bibitem{mori}  R. Zwanzig, in Lectures of  Theoretical Physics, Vol. III, 
               Wiley Interscience, New York (1961);
 H. Mori, Prog. Theor. Phys. {\bf 33}, 423; {\bf 34}, 399 (1965).              

\bibitem{comecalcologamma} 
The autocorrelation of
$f_{\bf k}(t)$, requested to evaluate  $\gamma_{\bf k}(t)$,  
is explicitely given by 
$f_{\bf k}(t)={\cal F}_{\bf k}(t)
+\int_0^t \gamma_{\bf k}(t-t') \partial_{t'} \rho_{\bf k}(t') dt' $ 
where ${\cal F}_{\bf k}(t)$ is given in Eq.\protect\ref{eq:residual}.
Since $f_{\bf k}(0)={\cal F}_{\bf k}(0)$, only 
two terms survive in $\gamma_{\bf k}(t)$:
$ \langle {\cal F}_{\bf -k}(0) {\cal F}_{\bf k}(t) \rangle $ and
$ \langle {\cal F}_{\bf -k}(0) \int_0^t \gamma_{\bf k}(t-t') 
\partial_{t'} \rho_{\bf k}(t') dt' \rangle$.


\bibitem{note}
Instead of using the explicit force implied by
Eqs. \ref{eq:residual},\ref{eq:expansion}, it is possible to use the
force implied by Eq. \ref{eq:langevin}. The result requires
time-correlations of $\ddot{\rho}_k$, which cannot
be carried out explicitly. Previously $\ddot{\rho}_k$ have been projected
onto density pairs and gaussian statistics is used to factorize the
4-point correlations\cite{gotzesjolander}, thus getting rid of the
bare potential. 




%




\bibitem{ybg-gaussian}
%
The Gaussian approximation for noise (and density) implies that, in equation 
(\protect\ref{eq:ybg}), all terms in the sum are neglected, except  
for ${\bf k}={\bf k'}$. 
Evidently this step could have  equally well  been represented 
as RPA, 
but in essence all follows stems logically 
and exactly from the choice 
of Gaussian noise.



\bibitem{pines} D. Pines, P. Nozi\'eres, The theory of quantum liquids, W.A.
Benjamin N.Y. (1966).


\bibitem{spiegazione} By multiplying both sides of Eq.\protect\ref{eq:trialeq}
by $\rho_{\bf k}(0)$ and  averaging,
an exact equation for $\langle \rho_k(t)\rho_{-k}(0) \rangle$ 
is found. Assuming $\rho_{k}(t)$ to 
be a gaussian process, 
 $\langle{\cal F}_{\bf k}(t)  \rho_{\bf -k}(0) \rangle=0 $. 





\bibitem{crisanti}
A. Crisanti and
H.-J. Sommers, Z. f|r Physik B {\bf 87}, 341 (1992).

\bibitem{ss} K. S. Singwi, M.P. Tosi, R.H. Land, A. Sj\"{o}lander, Phys. Rev.
{\bf 176}, 589 (1968).

\bibitem{expanded} E. Zaccarelli {\it et al.}, in progress (2000).

\bibitem{mctmolecole} R.Schilling and T.Scheidsteger, Phys. Rev. E
{\bf 56}, 2932 (1997). 

\bibitem{mctsiti}S.-H.Chong and F.Hirata, 
Phys. Rev. E {\bf 57}, 1691 (1998); Phys. Rev. E {\bf 58},
6188 (1998); Phys. Rev. E {\bf 58}, 7296 (1998).

\bibitem{franoschdumbell} 
T. Franosch, M. Fuchs, W. G\"{o}tze, M. R. Mayr and 
A. P. Singh, Phys. Rev. E {\bf 56} 5659 (1997).


\bibitem{mctpolymers}
M. Fuchs and K. S. Schweizer, Macromolecules {\bf 30}, 5133 (1997).




\end{references}
\end{document}